\documentclass[aps,prl,twocolumn,superscriptaddress]{quantumarticle}
\pdfoutput=1
\usepackage{amssymb}
\usepackage{amsmath}
\usepackage{amsfonts}
\usepackage[stable]{footmisc}
\usepackage{hyperref,blkcntrl,moredefs,relsize,attrib,tikz,booktabs,capt-of,float,varwidth,epsdice,graphicx,enumitem}

\newcommand{\qed}{\nobreak \ifvmode \relax \else
      \ifdim\lastskip<1.5em \hskip-\lastskip
      \hskip1.5em plus0em minus0.5em \fi \nobreak
      \vrule height0.75em width0.5em depth0.25em\fi}

\newcommand{\abs}[1]{\left| #1 \right|} % for absolute value
\newcommand{\ket}[1]{\left| #1 \right>} % for Dirac bras
\newcommand{\bra}[1]{\left< #1 \right|} % for Dirac kets
\newcommand{\braket}[2]{\langle #1 \vphantom{#2} | #2 \vphantom{#1} \rangle} % for Dirac brackets
\let\baraccent=\= % rename builtin command \= to \baraccent
\renewcommand{\=}[1]{\stackrel{#1}{=}} % for putting numbers above =

\begin{document}

\title{Global ontologies for relativistic quantum systems and quantum field theory}

\author{Ian T. Durham}
\email[]{idurham@anselm.edu}
\affiliation{Department of Physics, Saint Anselm College, Manchester, NH 03102}
\date{\today}

\begin{abstract}
Epistemic models of nature prove to be problematic in many settings, particularly in those in which measurement procedures are ill-defined. By contrast, in ontological models of nature, measurement results are independent of the procedure used to obtain them. Quantum mechanics, as a model of nature, is notoriously ambiguous in this regard. If we assume that all measurement results can be expressed in terms of pointer readings, then any useful ontology would need to unambiguously specify the positions of things. But the positions of pointers are ill-defined in many relativistic and cosmological settings. One potential solution to this problem presents itself in the solutions to the Wheeler-DeWitt equation as developed by Hartle and Hawking. In this article we introduce such a model in which these solutions to the Wheeler-DeWitt equation serve as the ontology of the model. We then show that, for any model that admits these solutions as beables, the global preservation of Lorentz invariance is incompatible with global determinism. However, we also show that an Everett-like interpretation that allows for all possible mappings of the matter field to the geometry and for all possible foliations of these mappings as being equally real, might solve this problem.
\end{abstract}

\maketitle

\section{Background}\label{intro}
Quantum mechanics is ostensibly a theory about the results of measurements that are performed `on' systems (objects). This act of measurement appears to presuppose that something or someone (subject) must be doing the measuring. That is, quantum mechanics inherently includes epistemic elements. But, as Bell pointed out, precisely where or when to draw a distinction between subject and object is not manifest in the theory itself thus rendering the theory's epistemic elements ambiguous~\cite{Bell:2004aa}. Bell suggests that good physical theories should be able to say something concrete about reality itself rather than about measurement procedures. This is particularly true in situations in which the very concept of measurement can be ill-defined.

In contrast to epistemic models of nature that often deal with measurement procedures, in ontological models, measurement results are independent of the procedure used to obtain them. Bell refers to these concrete, objective elements of reality as `\textit{be}ables' in contrast to the more ambiguous term `observables' since they are independent of observation~\cite{Bell:2004ab,Bell:2004ac}. Bell's aim was to unambiguously represent the `positions of things' which he takes as referring to the positions of instrument pointers and argues that any ontology must adequately be distillable to these positions~\cite{Bell:2004ac}. In other words, Bell assumes that the positions of instrument pointers have a direct, one-to-one correspondence with some element of physical reality. If an instrument pointer that measures some physical quantity $A$ gives a value of exactly $a$ then the value $a$ is an objective fact about the system under consideration. Thus knowledge of the exact position of an instrument pointer gives objectively true information about the universe. So, for example, if the instrument pointer corresponds to a voltage or current reading, the assumption is that this tells us something objectively true about an electric field. But unambiguously specifying the positions of multiple pointer readings, particularly in dynamical many-body systems and in highly-relativistic or cosmological settings, can prove difficult as Bell, himself, well knew.

Bell's original theory of beables relied on the notion of local causality. There have been a few attempts to develop fully relativistic theories of local beables~\cite{Durr:2004aa,Durr:2013aa,Tilloy:2017aa} including extending Bohmian mechanics to quantum gravity and cosmology~\cite{Callender:1994aa,Pinto-Neto:2018aa}. However, Healey has noted (and Bell, himself, also recognized) that quantum field theories do not conform to the principle of local causality even if their equations are relativistically covariant and their observable algebras satisfy a relativistically motivated microcausality condition~\cite{Healey:2014aa}. To the best of our knowledge, there have been only two attempts to develop a theory of beables for quantum field theories that allowed for the existence of \textit{non}-local beables as a way around this problem. One was first developed by Bell himself and has subsequently been improved upon~\cite{Roy:1990aa,Colin:2003aa,Colin:2004aa}. Bell's approach includes both local and non-local beables. A second approach employing exclusively non-local beables was put forward by Smolin~\cite{Smolin:2015aa}. 

In this article we review both Bell's and Smolin's proposals in Sections~\ref{bell} and~\ref{smolin} respectively, and note distinct problems in each that make them unsuitable for unambiguously specifying the positions of things in relativistic, field-theoretic, and cosmological settings. We then propose an alternative theory of \textit{global} beables in Section~\ref{universe} that is based on solutions to the Wheeler-DeWitt equation that includes a set of constraints  for the geometry, matter field, and action that would need to be satisfied in order for the model to be considered fully deterministic.

\section{Field theories and beables}\label{bell}
\begin{figure}
\begin{center}
\begin{tikzpicture}
%axes
\draw[-latex] (0,0) -- (0,5);
\node[left] at (0,5) {\footnotesize{$t$}};
\draw[-latex] (0,0) -- (4,0);
\node[below] at (4,0) {\footnotesize{$x$}};
%bounded region
\draw (2,4) circle (0.25cm);
\node at (2,4) {\footnotesize{$Q$}};
%light cone
\draw (1.75,4) -- (1,0.5);
\draw (2.25,4) -- (3,0.5);
\node at (2,2) {\footnotesize{$R$}};
%cross-section
\draw (1.1,1) -- (2.9,1);
\draw[dashed] (-0.05,1) -- (1.1,1);
\node at (1.75,0.35) {\footnotesize{$X$}};
\draw[thin,-latex] (1.8,0.5) -- (2,1);
\node[left] at (-0.05,1) {\footnotesize{$t_X$}};
\end{tikzpicture}
\caption{\label{localdet} Beables associated with the bounded region of spacetime $Q$ are determined by those associated with any temporal cross-section (i.e. spacelike slice) $X$ of region $R$ that fully encloses the past light-cone of $Q$.}
\end{center}
\end{figure}
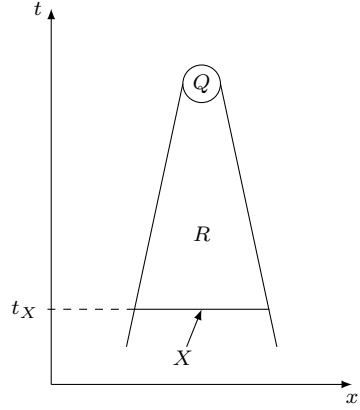

Local beables are specifically those beables that are confined to some specific bounded region of spacetime~\cite{Bell:1990aa,Bell:2004ab}. Specifically such beables must be associated with some bounded set $Q=\{\mathbf{x},t\}$ of spacetime coordinates where $Q$ is the subset of some metric space $(M,g)$ where $g$ is a metric on $M$. Specifically beables defined for $Q$ are assumed to be determined by those corresponding to any temporal cross section (i.e. spacelike slice) $X$ of the spacetime region $R$ that fully encloses the past light-cone of $Q$ where $X$ is a subset of $R$ corresponding to all possible values of $\mathbf{x}$ associated with the value $t_X$. This is shown for a single spatial dimension, $x$, in Fig.~\ref{localdet}.
Bell refers to this condition as `local determinism' and it serves as the foundation for his development of a local causal theory~\cite{Bell:2004ab}.

Fields, of course, are not strictly local since they exist in a \textit{region} of spacetime rather than at a single event. In other words, the spacetime region defined by $Q$ for a field consists of a \textit{range} of spacetime coordinate values, $\Delta x$ and $\Delta t$ as depicted in Fig.~\ref{localdet}. Unambiguously determining the position of something (e.g. a pointer) however, requires specifying its exact coordinates. As Bell notes, the obvious choice of a suitable beable here would appear to be the energy density $T_{00}(x)$. However the commutator $\left[T_{00}(x),T_{00}(y)\right]$ is not zero and thus the $T_{00}(x)$ do not have simultaneous eigenvalues for all $x$. One can immediately see the problem this poses for unambiguously specifying the positions of things. As such, since ordinary matter is composed of fermions and their locations are crucial for determining various macroscopic properties including the positions of instrument pointers, Bell's solution was to instead consider the lattice fermion particle density~\cite{Bell:2004ac}.

\subsection{Lattice fermion number density}\label{fermion}
For simplicity, we begin by replacing the three-space continuum with a dense lattice (which, in essence, quantizes space \textit{a priori}) while time remains continuous. The lattice points are given by $l=1,2,\ldots,L$ where $L$ is assumed to be very large. The lattice point fermion number operators can be defined as
\begin{equation}
\hat{F}(l)\equiv\hat{\psi}^{\dag}(l)\hat{\psi}(l)
\label{lattfermnumop}
\end{equation}
where we are assuming a summation over Dirac indices and all Dirac fields. Strictly speaking, this is a number \textit{density} and thus measures the number of fermions at a given lattice point~\cite{Colin:2004aa}. The corresponding eigenvalues are integers
\begin{equation}
F(l)=1,2,\ldots,4N
\label{fermnumeigval}
\end{equation}
where $N$ is the number of Dirac fields. That is, $F(l)$ labels a particular configuration of the Dirac fields at a given lattice point $l$. For example, for a single Dirac field, $N=1$ and thus $F(l)=1,2,3,4$ corresponding to the four possible solutions to the Dirac equation.  The fermion number operator for the entire lattice is simply the sum of the lattice point operators,
\begin{equation}
\hat{F}=\sum_l\hat{F}(l).
\label{fermnumop}
\end{equation}

The fermion number configuration of the world, i.e. precisely \textit{how} the fermions in the universe are arranged at any given instant of time $t$, is a list of such integers (one for each lattice point),
\begin{equation}
n(t) = [F(1),F(2),\ldots,F(L)]_t
\label{lattfermnum}
\end{equation}
where the subscript $t$ labels the set for a given time $t$. This is the lattice fermion number configuration and it serves as the local beable for a theory since it is associated with definite positions in space i.e. a definite configuration of all fermions in the universe at a time $t$~\cite{Bell:2004ac}. To this Bell also adds the state vector $\ket{t}$ as a \textit{non}-local beable. The complete specification of the universe at a time $t$ is then given by the set $T=\{n(t),\ket{t}\}$ where, in accordance with the theory of local beables, $n(t)$ is associated with some bounded set of spacetime coordinates $Q$. Since the configuration $n(t)$ is exact, there is no spread in the spacetime coordinates of any given fermion at time $t$, i.e. the position of each fermion is assumed to be known exactly. Thus $Q$ is \textit{local in its elements}, i.e. each element is perfectly localized.

\subsection{The Unruh effect}
In order to construct a fermion number operator we first briefly review the algebra of fermionic operators on an antisymmetric Fock space~\cite{Dirac:1927aa,Fock:1932aa}. It is easiest to do this using a momentum-space description rather than a lattice description. We define the fermion creation and annihilation operators by the following relations,
\begin{align}
\hat{f}^{\dag}_\mathbf{k}\ket{1}  & = 0, & \hat{f}^{\dag}_\mathbf{k}\ket{0} & = \ket{1},  \label{fermcreat} \\
\hat{f}_\mathbf{k}\ket{1} & = \ket{0}, & \hat{f}_\mathbf{k}\ket{0} & = 0, \label{fermanni}
\end{align}
where we note that~\eqref{fermcreat} follows from the fact that fermionic Fock states can contain a maximum of a single particle. In other words, the only two possible Fock states for fermions are $\ket{0}$ and $\ket{1}$. Unlike the bosonic operators, the fermionic operator algebra is defined by the \textit{anti}-commutators due to the antisymmetry of the space, i.e.
\begin{align}
\{\hat{f}_\mathbf{j},\hat{f}^{\dag}_\mathbf{k}\} & = \delta_{jk}, \nonumber \\
\{\hat{f}_\mathbf{j}),\hat{f}_\mathbf{k}\} = \{\hat{f}^{\dag}_\mathbf{j},\hat{f}^{\dag}_\mathbf{k}\} & = 0. \label{anticomm}
\end{align}
Multi-particle fermionic number states can then be constructed from the vacuum state as
\begin{equation}
\ket{n_0,n_1,\ldots} = (f^{\dag}_1)^{n_1}(f^{\dag}_0)^{n_0}\ket{0}.
\label{fermnumstate}
\end{equation}
Note that~\eqref{fermnumstate} does not represent the lattice fermion number density given in~\eqref{lattfermnum}. We are still working with a momentum-space description here.

We then define the fermion number operator as
\begin{equation}
\hat{\mathcal{F}}\equiv\int \textrm{d}^3 k\hat{f}^{\dag}_\mathbf{k}\hat{f}_\mathbf{k}.
\label{numop}
\end{equation}
We only obtain the discrete fermion number operator of~\eqref{fermnumop} in the lattice approximation, i.e.
\begin{equation}
\hat{\mathcal{F}} \rightarrow \hat{F} \quad\quad \text{(lattice approximation).}
\end{equation}

For simplicity, let us consider only the lowest eigenmode in the Minkowski vacuum $\ket{0_{\textrm{M}}}$. The expectation value of the fermion number operator gives the number of fermions that we would expect to see in the vacuum. For the lowest eigenmode this is
\begin{equation}
\bra{0_{\textrm{M}}}\hat{\mathcal{F}}\ket{0_{\textrm{M}}} =\bra{0_{\textrm{M}}}\hat{f}^{\dag}_0\hat{f}_0\ket{0_{\textrm{M}}} = 0
\label{expvalvac}
\end{equation}
which follows from~\eqref{fermanni}. In other words any inertial observer in the Minkowski vacuum would not expect to detect any fermions in the lowest eigenmode. 

Now consider an accelerated observer in this same Minkowski vacuum. We define coordinate transformations as
\begin{align}
t & =\pm\frac{e^{a\xi}}{a}\sinh(a\tau) \nonumber \\
x & =\pm\frac{e^{a\xi}}{a}\cosh(a\tau)
\label{rindler}
\end{align}
where $\tau$ and $\xi$ are the Rindler time and space respectively and $a$ is a parameter corresponding to the proper acceleration of an arbitrarily chosen reference trajectory. The signs require that we define two coordinate patches that we will denote I ($x>0$) and II ($x<0$). Within each coordinate patch we define creation and annihilation operators such that, for example, in region I we have $\hat{f}_{\mathbf{k},\textrm{I}}^{\dag}\ket{0_{\textrm{I}}}=\ket{1_{\textrm{I}}}$ and $\hat{f}_{\mathbf{k},\textrm{I}}\ket{0_{\textrm{I}}}=0$ where $\ket{0_{\textrm{I}}}$ is the vacuum state in that region (patch). The restriction of $\ket{0_{\textrm{M}}}$ to region I, however, is equivalent to a thermal state with temperature $\mathcal{T}=a/2\pi$~\cite{Fulling:1973aa,Davies:1975aa,Unruh:1976aa,Wald:1994aa}. Specifically, it can be shown that~\cite{Crispino:2008aa}
\begin{align}
\bra{0_{\textrm{M}}}\hat{\mathcal{F}}_\textrm{I}\ket{0_{\textrm{M}}} & = \bra{0_{\textrm{M}}}\hat{f}_{\mathbf{k},\textrm{I}}^{\dag}\hat{f}_{\mathbf{k},\textrm{I}}\ket{0_{\textrm{M}}} \nonumber \\
& \propto \frac{1}{e^{2\pi\abs{\mathbf{k}}/a}+1}.
\label{expvalrin}
\end{align}
This is known as the Unruh effect and it is clear that a non-inertial observer will see fermions in the lowest eigenmode when an inertial observer will not. This means that, in general, in the lattice description non-inertial observers will observe a different value for $n(t)$ than inertial observers and will thus have a different specification $T$ for the universe at some time $t$. This is clearly problematic for any theory of beables that seeks to unambiguously specify the position of something presumably composed of fermionic matter. We note that while it has been shown that massive Unruh particles cannot be directly observed~\cite{Kialka:2018aa}, it is presently unknown if their presence can be indirectly inferred. As such, the lattice fermion number density remains ill-suited as a beable for unambiguously specifying the positions of things in relativistic, field theoretic, and cosmological situations.

\section{Non-local beables}\label{smolin}
Smolin's recognition of the need for a theory of non-local beables was motivated by three observations~\cite{Smolin:2015aa}. First he notes that if the metric of spacetime is a quantum operator and thus subject to the usual quantum fluctuations, then locality is merely a classical approximation. Non-locality must arise from quantum fluctuations of the metric and there are some arguments that non-locality must be present in quantum gravity at large scales~\cite{Markopoulou:2007aa}. Second, he notes that, if space itself is not fundamental, i.e. is emergent---something most theories of quantum gravity agree on---then locality must also be emergent. As such, space and the quantum state emerge simultaneously, each carrying some information about the non-local ontology. 

\subsection{Theory of \textit{a}-local beables}\label{alocal}
Smolin notes that the only meaningful beables are those that describe relationships between elementary events or particles which is reconcilable with the primacy of pointer readings. So the hidden variables do not give a detailed description of the inner workings of, say, an electron, for example. Rather they describe the details of the relations between electrons and each other or between electrons and other fundamental entities in the universe that are ignored or not obvious when coarse-graining is applied. Since these beables are more fundamental than space itself, Smolin prefers the term \textit{a}-local to \textit{non}-local. This leads him to propose that the fundamental beables are relational and `a-local' and that their fundamental description must necessarily be in a phase from which space (and quantum theory, for that matter) has yet to emerge. In fact space and quantum theory are assumed to emerge at the same time. Smolin argues that the stochasticity of quantum theory arises from our lacking control over beables that describe relationships between a system and other, distant systems in the universe~\cite{Smolin:2015aa}. Smolin's proposal for \textit{a}-local beables, first fully described in~\cite{Smolin:2015aa}, is the culmination of the development of a broader dynamical theory of relational hidden variables~\cite{Smolin:2002aa,Starodubtsev:2003aa,Markopoulou:2004aa,Adler:2004aa}. While Smolin originally presented it as a specifically bosonic model in~\cite{Smolin:2002aa}, he expresses it in more general terms in where he outlines the full theory and first introduces the term `\textit{a}-local'~\cite{Smolin:2015aa}.

The beables of the theory are $d$, $N\times N$ real symmetric matrices $X^{j}_{ai}$ with $a = 1,\ldots,d$ and $i,j=1,\ldots,N$. The classical, local observables corresponding to such things as pointer readings on measurement devices, are taken to be the eigenvalues $\lambda_i^a$ of these matrices. In direct analogy to Bell's use of the lattice fermion number, these eigenvalues are taken as corresponding to the positions of $N$ particles in $d$ space. The dynamics of these matrices is given by the action
\begin{equation}
S = \mu\int dt\;\textrm{Tr}\left[\dot{X}^2_a-\omega^2\left[X_a,X_b\right]\left[X^a,X^b\right]\right].
\label{smoaction}
\end{equation}
The matrices $X^a$ are assumed to be dimensionless, $\omega$ is a frequency, and $\mu$ has units of mass$\cdot$length$^2$. As such the parameters of the theory define an energy, $\varepsilon = \mu\omega^2$. In this case, the $N$ particles are free. But classical interactions can be modeled by including a potential $V(\lambda)$ that is a function of the eigenvalues in the trace.

The theory is invariant under $SO(N)$ transformations
\begin{equation}
X^a \to UX^aU^T
\label{gaugetrans}
\end{equation}
where $U\in SO(N)$, and thus they constitute the gauge transformations of the theory. As such the physical observables, corresponding to the $\lambda^a_i$, are invariant under $SO(N)$ transformations. The off-diagonal elements of $X^a$ are the non-local hidden variables of the system. The model includes a translation symmetry that ensures that the center of mass momentum of the system is conserved and defines the potential energy in a manner such that it has its minima whenever the $X^a$ commute with one another in which case they can be simultaneously diagonalized. This is precisely what gives the classical approximation and leads to the interpretation of the eigenvalues as labeling the positions of $N$ identical particles in $\mathbb{R}^d$.

\subsection{Noncommutativity}
The eigenvalues of the $X^a$ correspond to the positions of $N$ objects in $d$ space. So each individual $X^a$ can be thought of as a single configuration of these particles in that space akin to Bell's lattice fermion number density (configuration). Smolin notes that these eigenvalues are specifically the classical, local observables~\cite{Smolin:2015aa} and should, thus, be distillable to pointer readings.

But, as Bell clearly notes, observables that do not all have simultaneous eigenvalues, i.e. that do not all commute, cannot be promoted to the status of beables~\cite{Bell:2004ac}. For example, in~\cite{Bell:2004ac} he expressly \textit{rejects} energy density $T_{00}(x)$ as a choice of beable precisely for this reason. As he notes, the lack of simultaneous eigenvalues for all positions means we would need to devise some new manner of specifying a joint probability distribution for any pair of such observables. Recall that the purpose of a beable is to be able to say what \textit{is} rather than what is merely \textit{observed}. The idea is to be able to `uncover' the hidden variables, ultimately allowing us to say something for certain about the actual state of a system or of the universe as a whole at a given time. But the potential presence of an non-commuting position observables precludes us from doing this. In effect, since the eigenvalues of $X^a$ are associated with such things as pointer positions, anytime a pair of $X^a$ do not commute, there must exist two such things (e.g. two pointer positions) that cannot be simultaneously known to perfect precision which violates the very essence of a beable as set forth by Bell.

We might be inclined to assume that all of the $X^a$ mutually commute. Certainly if Smolin's theory is entirely classical, they should. Likewise, in orthodox quantum mechanics, position operators mutually commute. However, it is worth noting that Smolin's action integral given by~\eqref{smoaction} includes a product of two commutators of the $X^a$. If Smolin's theory was entirely classical or equivalent to orthodox quantum mechanics, there would be no reason for these terms in the action integral to exist since they would trivially be zero. In other words, if the $X^a$ were all to mutually commute, then equation~\eqref{smoaction} should trivially reduce to
\begin{equation}
S = \mu\int dt\;\textrm{Tr}\left[\dot{X}^2_a\right].
\label{smoaction}
\end{equation}
Since Smolin includes the term $\omega^2\left[X_a,X_b\right]\left[X^a,X^b\right]$ in the action integral, we can assume that some of the $X^a$ do not commute in which case Smolin's theory is unsuitable for unambiguously specifying the positions of things.

\section{Theory of global beables}\label{universe}
We now describe a theory of global beables based on solutions to the Wheeler-DeWitt equation that does not suffer from the same problems as the theories discussed in the previous sections. We choose the word `global' since possible solutions to the Wheeler-DeWitt equation include wave functions (or, more properly, functionals) of the universe. We begin by briefly reviewing the Wheeler-DeWitt formalism before introducing the theory.

\subsection{Wheeler-DeWitt formalism}\label{wheeler}
If we take spacetime as being foliated into spacelike submanifolds, we can decompose the metric tensor as
\begin{align}
g_{\mu\nu}dx^{\mu}dx^{\nu} & =\left(-\alpha^2 + \beta_k\beta^k\right)dt^2 \nonumber \\
& \quad\quad\quad + 2\beta_k dx^k dt + \gamma_{ij}dx^i dx^j
\label{metric}
\end{align}
where $\alpha$ is the lapse function\footnote{We use $\alpha$ for the lapse function and $\beta$ for the shift functions instead of the usual $N$ in order to distinguish these from the $N$ labeling the matrices in Smolin's theory.}, the $\beta_k$ are the shift functions, and $\gamma_{ij}$ is the spatial three-metric. The usual summation convention is assumed such that Greek indices range from $0\to 3$ while Latin indices range from $1\to 3$.  The lapse function is given as $\alpha = \left( -g^{00}\right)^{-1/2}$ where the $g^{00}$ is the usual four-dimensional value. The shift functions are thus given as $\beta_k = g_{0i}$ where, again, these are elements of the usual four-dimensional metric. The spatial three-metric is therefore $\gamma_{ij} = g_{ij}$. In the Hamiltonian formulation that follows, the spatial three-metric serves as the set of generalized coordinates to which we can associate conjugate momenta $\pi^{ij}$. We define $\mathcal{R}=\;^{(3)}R$ where $^{(3)}R$ is the three-dimensional Ricci scalar. The Hamiltonian is then a constraint given by
\begin{equation}
\mathcal{H} = \frac{1}{2\sqrt{\gamma}}G_{ijkl}\pi^{ij}\pi^{kl} - \sqrt{\gamma}\mathcal{R} = 0
\label{hamconst}
\end{equation}
where $\gamma = \det{(\gamma_{ij})}$ and $G_{ijkl}=(\gamma_{ik}\gamma_{jl} + \gamma_{il}\gamma_{jk} - \gamma_{ij}\gamma_{kl})$ is the Wheeler-DeWitt metric on superspace which is the space of all three-geometries. We can quantize this by employing the ADM formalism~\cite{Arnowitt:1959aa} which allows us to turn the momenta and field variables into operators such that the Hamiltonian operator becomes
\begin{equation}
\hat{\mathcal{H}}=\frac{1}{2\sqrt{\gamma}}\hat{G}_{ijkl}\hat{\pi}^{ij}\hat{\pi}^{kl}-\sqrt{\gamma}\hat{\mathcal{R}}
\label{hamquant}
\end{equation}
where, in position space, the generalized coordinates and their conjugate momenta are
\begin{align}
\hat{\gamma}_{ij}(t,x^k) & \to \gamma_{ij}(t,x^k);\quad\textrm{and} \nonumber \\
\hat{\pi}^{ij}(t,x^k) & \to -i\frac{\delta}{\delta\gamma_{ij}(t,x^k)}
\label{momcoord}
\end{align}
respectively. The Hamiltonian is not, however, applied to the usual wave function. Instead it is applied to a wave \textit{functional} $\Psi(\gamma)$ of field configurations defined on the spatial three-metric. The Hamiltonian constraint~\eqref{hamconst} necessarily implies, then, that
\begin{equation}
\hat{\mathcal{H}}\Psi(\gamma)=0
\label{wheelerdewitt1}
\end{equation} 
or, more familiarly 
\begin{equation}
\hat{H}(x)\ket{\psi}=0
\label{wheelerdewitt2}
\end{equation}
which is known as the Wheeler-DeWitt equation~\cite{DeWitt:1964aa,DeWitt:1967aa}. Here $\hat{H}(x)$ is a Hamiltonian constraint, of which there are technically an infinite number, and $\ket{\psi}$ is referred to as the wave function of the universe even though it is more properly a state vector. Since $\Psi(\gamma)$ is a functional of the field configurations defined on the spatial three-metric alone, the Hamiltonian no longer determines time evolution and thus the usual Schr\"{o}dinger equation does not apply. Specifically $\Psi(\gamma)$ contains all of the information about the matter and geometry content of the universe for a given hypersurface defined by the three-metric~\cite{Hartle:1983aa}.

The Hamiltonian in the Wheeler-DeWitt equation, unlike the case in typical quantum field theory or quantum mechanics, is a first class constraint on physical states which is a dynamical quantity in a constrained Hamiltonian system whose Poisson bracket with all other constraints must vanish on the constraint surface in phase space~\cite{Dirac:1950aa}. For example, due to the invariance of the wave functional under spatial diffeomorphism, the Wheeler-DeWitt equation is typically accompanied by a momentum constraint, $\hat{P}(x)\ket{\psi}=0$. Thus, because the Hamiltonian is a first class constraint, $\{\hat{P}(x),\hat{H}(x)\}=0$.

\subsection{Global beables}
We wish to construct a theory of \textit{global} beables guided by the Wheeler-DeWitt equation that does not suffer the same fate as Bell's and Smolin's theories but that still endeavors to make the concept `positions of things' more precise. Since we technically have an infinite number of Hamiltonian constraints due to the infinite degrees of freedom of the phase space, we do not appear to be much closer to a useful and unambiguous definition of the `positions of things' in real, physical space. But, as it turns out, we can reduce the number of Hamiltonian constraints to just one by making a minisuperspace approximation~\cite{DeWitt:1967aa,Hartle:1983aa,Vilenkin:1994aa}. 

In any dynamical model, positions change over time. But in the model we will consider here, time is simply a manner by which we can order the spacelike submanifolds, i.e. the hypersurfaces defined by the three-metric. In other words, four-dimensional spacetime is foliated into three-dimensional spacelike surfaces ordered over the fourth dimension (time). For our model, we assume that each spacelike surface is a specific closed three-sphere on which the matter field is fixed. Transitions from one three-sphere to another take the place of time evolution in this model. For example, suppose that the total classical action for some particular metric $g$ coupled to a scalar field $\phi$ is $S[g,\phi]$. The quantum-mechanical amplitude for the occurrence of a particular spacetime and thus field history is $\exp(iS[g,\phi])$. In analogy to the usual propagator $\braket{x^{\prime\prime},t^{\prime\prime}}{x^{\prime},t^{\prime}}$ and recalling that the spatial three-metric is $\gamma_{ij}=g_{ij}$, the transition amplitude between any pair of three-spheres is~\cite{Hartle:1983aa}
\begin{equation}
\braket{\gamma_{ij}^{\prime\prime},\phi^{\prime\prime}}{\gamma_{ij}^{\prime},\phi^{\prime}}=\int \delta g \delta\phi \;e^{iS[g,\phi]}
\label{transamp}
\end{equation}
where the integral is over all four-geometries and field configurations that match the given values on the two three-spheres. In Smolin's model, evolution occurs explicitly according to the Schr\"{o}dinger equation. But in this model, there is no evolution, strictly speaking. Rather~\eqref{transamp} more correctly describes the amplitude for a certain three-geometry and an associated field to be fixed on any pair of three-spheres. The wave functionals are then defined as
\begin{equation}
\Psi[\gamma_{ij},\phi]=\int_C \delta g \delta \phi \; e^{iS[g,\phi]}
\label{wavefunc}
\end{equation}
where the integral is over a class $C$ of spacetimes with a compact boundary on which $\gamma_{ij}$ is the induced metric and $\phi$ is the field configuration on that boundary. One can also show that because we expect in gravity to find the field equations satisfied as identities, then
\begin{equation}
\int \delta g\delta\phi\;\hat{H}(x)\;e^{iS[g,\phi]}=0
\label{extendhamconst}
\end{equation}
for any class of geometries summed over and for any intermediate three-sphere on which $\hat{H}(x)$ is evaluated. In order to specify a particular state of the universe, the details of the class $C$ must be specified. Therefore, if the universe is in a quantum state specified by a particular state vector and corresponding wave function, then that wave function describes the correlations between observables to be expected in that state. As such, like Hartle and Hawking, we restrict the geometrical degrees of freedom to spatially homogenous, isotropic, closed universes with $S^3$ topology (i.e. closed three-spheres) and the matter degrees of freedom to a single, homogenous, conformally invariant scalar field with the cosmological constant assumed to be positive. Our aim is to consider what it means to specify the `positions of things' in such a model and we refer the reader to~\cite{Hartle:1983aa} for the full mathematical formalism of the minisuperspace model.

In order to specify the `positions of things' we first must specify both what we mean by `position' and what we mean by `thing.' Since Bell's aim in developing the concept of beables was to account for positions of such classical things as instrument pointers, we will assume that Bell was referring to things that were constructed of ordinary matter. As such, the `things' in this model are represented by the matter field $\phi$. In Bell's fermionic model, the three-space continuum was replaced by a dense but discrete lattice. In the minisuperspace model, however, we retain the three-space continuum and its form is specified by the induced metric $\gamma_{ij}$ where the induced metric takes the place of the discrete lattice. Thus, specifying the `positions of things' entails mapping the matter field \textit{onto} the metric which amounts to jointly specifying a field configuration and a metric (i.e. a three-sphere) via a state vector $\ket{\gamma_{ij},\phi}$ or its associated wave functional $\Psi(\gamma_{ij},\phi)$. Thus, if the beables must specify the positions of things, then jointly specifying the field configuration (`things') and the induced metric on which the field is specified (`positions') accomplishes this task. Thus the state vector $\ket{\gamma_{ij},\phi}$ or its associated wave functional $\Psi(\gamma_{ij},\phi)$ is a beable. \textit{The set of all such state vectors or wave functionals, which are operationally equivalent, are the beables of our theory.}

It is worth comparing $\ket{\gamma_{ij},\phi}$ to Bell's lattice fermion number density, $n(t)$, and Smolin's symmetric matrices,$X^{\alpha}$. In Bell's model, the $n(t)$ ostensibly give the positions of things such as pointers or anything else consisting of ordinary matter in an unambiguous manner. But as we have shown, it is unsuitable as a beable due to the fact that non-inertial observers will find a different value for $n(t)$ than inertial observers. Conversely, in Smolin's theory, the beables are a set of real, symmetric matrices $X^{\alpha}$ whose eigenvalues correspond to the positions of things. However, the $X^{\alpha}$ do not commute.

By contrast, in our theory, the beables are the set of all state vectors or wave functionals that are solutions to the Wheeler-DeWitt equation. Since these solutions contain the complete distribution of the matter field for a given three-sphere along with the geometry of space itself (something Bell's theory does not include), they unambiguously specify the position of every bit of matter on a given three-sphere without any need for measurement. There simply is no measurement problem since there is no measurement being made. The state of the universe for a given three-sphere is fully specified.

On the other hand, this does assume that there is only one correct mapping of the matter field to each three-sphere and does not explicitly take into account any backreaction. One way to circumvent this problem is to adopt a view similar to Everett's such that \textit{all} possible mappings are equally correct and, in some sense, real. This is not quite the same thing as Everett's original approach, however, which was actually based on solutions to the Schr\"{o}dinger equation and thus explicitly included time evolution~\cite{Everett:1956aa,Everett:1957aa,DeWitt:1968aa,DeWitt:1970aa,DeWitt:1972aa}. The two approaches are only approximately equivalent when the Wheeler-DeWitt equation is applied to small subsystems through a suitable WKB approximation within the minsupserspace model, in which case the Schr\"{o}dinger equation is obtained~\cite{Vilenkin:1989aa}. Nevertheless, taking all possible mappings as equally correct is very much in the spirit of Everett and can be interpreted in a similar manner (i.e. as a collection of universes).

\subsection{Deterministic constraints}
Nothing in the above suggests that there is any preferred foliation. If we were to limit ourselves to a single subset of possible mappings of the matter field to the geometry (i.e. a single universe within an Everett-like model) we still would not necessarily have a preferred foliation. A preferred foliation implies that the subset of three-spheres be ordered in some way. The only way to force order on a set of states in this model is to require that the transition amplitude between any pair of three-spheres be either unity or zero---such a transition is either allowed or it is not. In other words, for a subset of allowed three-spheres to be ordered the transition amplitude between any pair is either allowed, if they are adjacent in the order, or not allowed. This can be expressed in the form of an added constraint on~\eqref{transamp},
\begin{equation}
\int \delta g\delta\phi\;e^{iS[g,\phi]}\;\in\;\{0,1\} \quad \textrm{for all} \quad \ket{\gamma_{ij},\phi}.
\label{dertreq}
\end{equation}
It is worth noting, here, that we are assuming that the Wheeler-DeWitt approach is fundamentally a path integral approach. The implications of such an assumption are discussed in~\cite{Kiefer:1991aa}. 

\begin{figure}
\begin{center}
\begin{tikzpicture}
%axes
\draw[-latex] (0,0) -- (0,5);
\node[left] at (0,5) {\footnotesize{$t$}};
\draw[-latex] (0,0) -- (4,0);
\node[below] at (4,0) {\footnotesize{$x$}};
%bounded region
\draw (2,4) circle (0.25cm);
\node at (2,4) {\footnotesize{$Q$}};
%light cone
\draw (1.75,4) -- (1,0.5);
\draw (2.25,4) -- (3,0.5);
\node at (2,3) {\footnotesize{$R$}};
%cross-section one
\draw (1.1,1) -- (2.9,1);
\draw[dashed] (-0.05,1) -- (1.1,1);
\node at (1.75,0.35) {\footnotesize{$X$}};
\draw[thin,-latex] (1.8,0.5) -- (2,1);
\node[left] at (-0.05,1) {\footnotesize{$t_X$}};
%cross-section two
\draw (1.25,2) -- (2.675,2);
\draw[dashed] (-0.05,2) -- (1.25,2);
\node[left] at (-0.05,2) {\footnotesize{$t_{X^{\prime}}$}};
\draw[thin,-latex] (2.5,0.5) -- (2,2);
\node at (2.5,0.35) {\footnotesize{$X^{\prime}$}};
\end{tikzpicture}
\caption{\label{localdet2} Beables associated with the bounded region of spacetime $Q$ are determined by those associated with any temporal cross-section (i.e. spacelike slice) $X$ of region $R$ that fully encloses the past light-cone of $Q$. Likewise, they are determined with any spacelike slice $X^{\prime}$ that also fully encloses the past light-cone of $Q$.}
\end{center}
\end{figure}
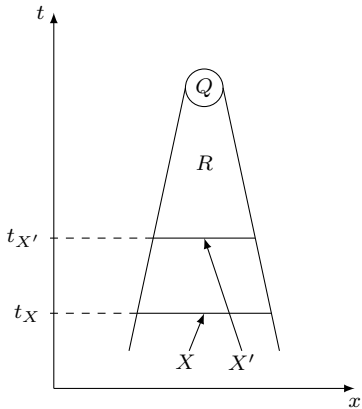

Recall that in the theory of \textit{local} beables, beables defined on some bounded set $Q=\{\mathbf{x},t\}$ of spacetime coordinates were determined by those corresponding to any spacelike slice $X$ of the spacetime region $R$ that fully encloses the past light-cone of $Q$ (see Section~\ref{bell} and Fig.~\ref{localdet}). Bell referred to this condition as \textit{local determinism}. This suggests that determinism inherently includes some idea of order in the sense that time is a label of spacelike slices such that those slices must occur in a given order for the theory to be considered deterministic. For example, consider a set of beables defined on some bounded set $Q=\{\mathbf{x},t\}$ of spacetime coordinates where the beables are assumed to be determined by those corresponding to any temporal cross section (i.e. spacelike slice) $X$ of the spacetime region $R$ that fully encloses the past light-cone of $Q$ where $X$ is a subset of $R$ corresponding to all possible values of $\mathbf{x}$ associated with the value $t_X$. The beables defined on $Q$ are, of course, also assumed to be determined by those corresponding to a different spacelike slice $X^{\prime}$ that is also a subset of $R$ as shown for a single spatial dimension, $x$, in Fig.~\ref{localdet2}.
This follows from the fact that where we drew $X$ in the first place was simply restricted to intersecting the past light-cone of $Q$. Exactly \textit{where} it intersected the past light-cone of $Q$ was not specified. Thus the beables defined on $Q$ are determined both by those defined on $X$ as well as those defined on $X^{\prime}$. But notice that any beables defined on the spacelike slice $X^{\prime}$ are \textit{also} determined by those associated with $X$ since $X$ intersects the past light-cone of every point on $X^{\prime}$. In other words, beables associated with $X$ determine both the beables associated with $X^{\prime}$ as well as those associated with $Q$. But the converse is not true, of course. Thus any ordering of the spacelike slices is necessarily deterministic. In fact the ordering of the set of all such spacelike slices \textit{defines} the past and future light-cones of $Q$.

In a global theory of beables based on the Wheeler-DeWitt model there is no `past' light-cone, strictly speaking, since time is treated as a label that we use to order the three-spheres. Each three-sphere is technically a set of \textit{all} spatial coordinates along with a metric. The metric defines the relations between the spatial coordinates, i.e. the geometry, and thus differentiates one three-sphere from another. Each three-sphere is thus a global representation of a spacelike slice in the local theory. This suggests, then, that in analogy with local determinism, \textit{any ordering of three-spheres must be deterministic}.

It is worth emphasizing the implications of this. As we have shown, a preferred foliation of the three-spheres implies that the theory is deterministic. In canonical gravity, a local time direction, which corresponds to an ordering of the spacelike slices as we have noted, is not compatible with local Lorentz invariance. One assumes that this holds globally as well. If that is true, then it would appear that in any theory of the universe that is describable by state vectors or wave functionals on all of space, Lorentz invariance is incompatible with global determinism. This does not suggest that we cannot find a suitable beable for unambiguously determining the positions of things. It merely suggests that we cannot do so within the confines of a globally deterministic system. The only other solution to this problem \textit{might} be to assume that all possible foliations exist. This would be in addition to assuming that all possible mappings of the matter field to the geometry exist. As such, there would have to be some duplication of three-spheres, e.g. a particular pair of three-spheres might exist in a certain order in one foliation but in another order in a different foliation. Allowing for all possible foliations, then, might allow for the compatibility of Lorentz covariance and determinism.

\section{Analysis}\label{conclusion}
We have proposed a theory of global beables---objectively known properties of the universe---aimed at making the concept `positions of things' more precise in a universal sense. Our theory, like Bell's and Smolin's, circumvents the problem of local causality pointed out by Healey~\cite{Healey:2014aa}. It does not suffer from the problem of the Unruh effect that proves  problematic for Bell's theory, or the problem of non-commutativity that proves problematic for Smolin's theory, but at the expense of introducing a preferred foliation of space-like three-spheres. On the other hand, it also makes it clear that any theory that employs a universal state vector or wave functional as a beable cannot simultaneously be both Lorentz covariant and deterministic. In addition, the Hartle-Hawking minisuperspace model includes situations in which the universe can quantum mechanically tunnel between two states which also could potentially be a problem for determinism. As such the problems with determinism may be related to the minisuperspace approximation itself. Of course, we could, in theory, choose to abandon the idea of a global Lorentz covariance altogether, or we could just as easily choose to abandon determinism. This problem might be circumvented if we choose to assume that all possible foliations exist, but this is at the expense of duplicating three-spheres and requires an Everett-like interpretation. On the other hand, it may be the case that spacetime really \textit{is} quantizable and discrete in which case an entirely different approach would be required. Nevertheless, it is instructive to elucidate this connection and, in doing so, we perhaps come a bit closer to understanding just what it means to unambiguously specify the positions of things in the manner of Bell.

\begin{acknowledgements}
We thank Alexander Smith, Andrzrej Dragan, Magdalena Zych, and Maaneli Derakhshani for detailed comments and Caslav Brukner for an opportunity to present this work at IQOQI in Vienna. We also thank two anonymous reviewers for detailed and helpful comments. This work was supported by Grant No. FQXi-MGB-1623/SVCF-2016-165593 from FQXi and the Silicon Valley Community Fund.
\end{acknowledgements} 

\bibliography{Beables.bib}

\begin{thebibliography}{40}
\providecommand{\natexlab}[1]{#1}
\providecommand{\url}[1]{\texttt{#1}}
\expandafter\ifx\csname urlstyle\endcsname\relax
  \providecommand{\doi}[1]{doi: #1}\else
  \providecommand{\doi}{doi: \begingroup \urlstyle{rm}\Url}\fi

\bibitem[Adler(2004)]{Adler:2004aa}
Stephen Adler.
\newblock \emph{Quantum theory as an emergent phenomenon}.
\newblock Cambridge University Press, Cambridge, 2004.
\newblock \doi{https://doi.org/10.1017/CBO9780511535277}.

\bibitem[Arnowitt et~al.(1959)Arnowitt, Deser, and Misner]{Arnowitt:1959aa}
Ronald Arnowitt, Stanley Deser, and Charles Misner.
\newblock Dynamical {S}tructure and {D}efinition of {E}nergy in {G}eneral
  {R}elativity.
\newblock \emph{Physical Review}, 116\penalty0 (5):\penalty0 1322--1330, 1959.
\newblock \doi{https://doi.org/10.1103/PhysRev.116.1322}.

\bibitem[Bell(1973)]{Bell:2004aa}
John~S. Bell.
\newblock Subject and object.
\newblock In D.~Reidel, editor, \emph{The Physicist's Conception of Nature},
  pages 687--690. Dordrecht-Holland, 1973.
\newblock \doi{https://doi.org/10.1142/9789812386540_0006}.

\bibitem[Bell(1976)]{Bell:2004ab}
John~S. Bell.
\newblock The theory of local beables.
\newblock \emph{Epistemological Letters}, March 1976.
\newblock \doi{https://doi.org/10.1142/9789812386540_0008}.

\bibitem[Bell(1990)]{Bell:1990aa}
John~S. Bell.
\newblock La nouvelle cuisine.
\newblock In A.~Sarlemijn and P.~Kroes, editors, \emph{Between {S}cience and
  {T}echnology}. Elsevier, Amsterdam, 1990.
\newblock \doi{https://doi.org/10.1016/B978-0-444-88659-0.50010-7}.

\bibitem[Bell(2004)]{Bell:2004ac}
John~S. Bell.
\newblock Beables for quantum field theory.
\newblock In \emph{Speakble and {U}nspeakable in {Q}uantum {M}echanics:
  {C}ollected {P}apers on {Q}uantum {P}hilosophy}. Cambridge University Press,
  Cambridge, 2004.
\newblock \doi{https://doi.org/10.1017/CBO9780511815676.021}.

\bibitem[Callender and Weingard(1994)]{Callender:1994aa}
Craig Callender and Robert Weingard.
\newblock The {B}ohmian {M}odel of {Q}uantum {C}osmology.
\newblock \emph{PSA: Proceedings of the Biennial Meeting of the Philosophy of
  Science Association}, 1994, Volume One: Contributed Papers:\penalty0
  218--227, 1994.
\newblock \doi{https://doi.org/10.1086/psaprocbienmeetp.1994.1.193027}.

\bibitem[Colin(2003)]{Colin:2003aa}
Samuel Colin.
\newblock A deterministic bell model.
\newblock \emph{Physics Letters A}, 317:\penalty0 349--358, 2003.
\newblock \doi{https://doi.org/10.1016/j.physleta.2003.09.006}.

\bibitem[Colin(2004)]{Colin:2004aa}
Samuel Colin.
\newblock Beables for {Q}uantum {E}lectrodynamics.
\newblock \emph{Annales de la Fondation Louis de Broglie}, 29:\penalty0
  273--296, 2004.

\bibitem[Crispino et~al.(2008)Crispino, Higuchi, and Matsas]{Crispino:2008aa}
Lu\'{i}s~C.B. Crispino, Atsushi Higuchi, and George~E.A. Matsas.
\newblock The {U}nruh effect and its applications.
\newblock \emph{Reviews of Modern Physics}, 90:\penalty0 787, 2008.
\newblock \doi{https://doi.org/10.1103/RevModPhys.80.787}.

\bibitem[Davies(1975)]{Davies:1975aa}
P.C.W. Davies.
\newblock {S}calar production in {S}chwarzschild and {R}indler metrics.
\newblock \emph{Journal of Physics A}, 8:\penalty0 609, 1975.
\newblock \doi{https://doi.org/10.1088/0305-4470/8/4/022}.

\bibitem[DeWitt(1964)]{DeWitt:1964aa}
Bryce~S. DeWitt.
\newblock Dynamical {T}heory of {G}roups and {F}ields.
\newblock In Bryce~S. DeWitt and Cecille DeWitt, editors, \emph{Relativity,
  {G}roups and {T}opology}. Gordon and Breach, New York, 1964.

\bibitem[DeWitt(1967)]{DeWitt:1967aa}
Bryce~S. DeWitt.
\newblock Quantum {T}heory of {G}ravity. {I}. {T}he {C}anonical {T}heory.
\newblock \emph{Physical Review}, 160:\penalty0 1113--1148, 1967.
\newblock \doi{https://doi.org/10.1103/PhysRev.160.1113}.

\bibitem[DeWitt(1968)]{DeWitt:1968aa}
Bryce~S. DeWitt.
\newblock The {E}verett-{W}heeler interpretation of quantum mechanics.
\newblock In Cecile~Morette DeWitt and John~Archibald Wheeler, editors,
  \emph{Battelle Rencontres: 1967 Lectures in Mathematics and Physics}, pages
  318--332, New York, 1968. Benjamin.

\bibitem[DeWitt(1970)]{DeWitt:1970aa}
Bryce~S. DeWitt.
\newblock Quantum {M}echanics and {R}eality: {C}ould the solution to the
  dilemma of indeterminism be a universe in which all possible outcomes of an
  experiment actuallly occur?
\newblock \emph{Physics Today}, 23\penalty0 (9):\penalty0 30--40, 1970.
\newblock \doi{https://doi.org/10.1063/1.3022331}.

\bibitem[DeWitt(1972)]{DeWitt:1972aa}
Bryce~S. DeWitt.
\newblock The {M}any-{U}niverses {I}nterpretation of {Q}uantum {M}echanics.
\newblock In Bernard D'Espagnat, editor, \emph{{F}oundations of {Q}uantum
  {M}echanics}, Proceedings of the {I}nternational {S}chool of {P}hysics
  ``{E}nrico {F}ermi'', Course 49, New York, 1972. Academic Press.

\bibitem[Dirac(1927)]{Dirac:1927aa}
Paul Dirac.
\newblock The {Q}uantum {T}heory of the {E}mission and {A}bsorption of
  {R}adiation.
\newblock \emph{Proceedings of the Royal Society A: Mathematical, Physical, and
  Engineering Sciences}, 114:\penalty0 243, 1927.
\newblock \doi{https://doi.org/10.1098/rspa.1927.0039}.

\bibitem[Dirac(1950)]{Dirac:1950aa}
Paul Dirac.
\newblock Generalized {H}amiltonian dynamics.
\newblock \emph{Canadian Journal of Mathematics}, 2:\penalty0 129--148, 1950.
\newblock \doi{https://doi.org/10.4153/CJM-1950-012-1}.

\bibitem[D\"{u}rr et~al.(2004)D\"{u}rr, Goldstein, Tumulka, and
  Zanghi]{Durr:2004aa}
Detlef D\"{u}rr, Sheldon Goldstein, Roderich Tumulka, and Nino Zanghi.
\newblock Bohmian {M}echanics and {Q}uantum {F}ield {T}heory.
\newblock \emph{Physical Review Letters}, 93:\penalty0 090402, 2004.
\newblock \doi{https://doi.org/10.1103/PhysRevLett.93.090402}.

\bibitem[D\"{u}rr et~al.(2013)D\"{u}rr, Goldstein, Norsen, Struyve, and
  Zanghi]{Durr:2013aa}
Detlef D\"{u}rr, Sheldon Goldstein, Travis Norsen, Ward Struyve, and Nino
  Zanghi.
\newblock Can {B}ohmian mechanics be made relativistic?
\newblock \emph{Proceedings of the Royal Society A: Mathematical, Physical, and
  Engineering Sciences}, 470:\penalty0 2162, 2013.
\newblock \doi{https://doi.org/10.1098/rspa.2013.0699}.

\bibitem[Everett(1956)]{Everett:1956aa}
Hugh Everett.
\newblock \emph{Theory of the Universal Wavefunction}.
\newblock PhD thesis, Princetone University, Princeton, 1956.

\bibitem[Everett(1957)]{Everett:1957aa}
Hugh Everett.
\newblock Relative {S}tate {F}ormulation of {Q}uantum {M}echanics.
\newblock \emph{Reviews of Modern Physics}, 29\penalty0 (3):\penalty0 454--462,
  1957.
\newblock \doi{https://doi.org/10.1103/RevModPhys.29.454}.

\bibitem[Fock(1932)]{Fock:1932aa}
Vladimir Fock.
\newblock Konfigurationsraum und zweite {Q}uantelung.
\newblock \emph{Zeitschrift f\"{u} Physik}, 75:\penalty0 622--647, 1932.
\newblock \doi{https://doi.org/10.1007/BF01344458}.

\bibitem[Fulling(1973)]{Fulling:1973aa}
S.A. Fulling.
\newblock Nonuniqueness of {C}anonical {F}ield {Q}uantization in {R}iemannian
  {S}pace-{T}ime.
\newblock \emph{Physical Review D}, 7:\penalty0 2850, 1973.
\newblock \doi{https://doi.org/10.1103/PhysRevD.7.2850}.

\bibitem[Hartle and Hawking(1983)]{Hartle:1983aa}
James~B. Hartle and Stephen~W. Hawking.
\newblock Wave function of the {U}niverse.
\newblock \emph{Physical Review D}, 28:\penalty0 2960--2975, 1983.
\newblock \doi{https://doi.org/10.1103/PhysRevD.28.2960}.

\bibitem[Healey(2014)]{Healey:2014aa}
Richard~A. Healey.
\newblock Causality and chance in relativistic quantum field theories.
\newblock \emph{Studies in History and Philosophy of Science Part B: Students
  in History and Philosophy of Modern Physics}, 48:\penalty0 156--167, 2014.
\newblock \doi{https://doi.org/10.1016/j.shpsb.2014.03.002}.

\bibitem[Kia{\l}ka et~al.(2018)Kia{\l}ka, Smith, Ahmadi, and
  Dragan]{Kialka:2018aa}
Filip Kia{\l}ka, Alexander~R.H. Smith, Mehdi Ahmadi, and Andrzej Dragan.
\newblock Massive {U}nruh particles cannot be directly observed.
\newblock \emph{Physical Review D}, 97:\penalty0 065010, 2018.
\newblock \doi{https://doi.org/10.1103/PhysRevD.97.065010}.

\bibitem[Kiefer(1991)]{Kiefer:1991aa}
Claus Kiefer.
\newblock On the meaning of path integrals in quantum cosmology.
\newblock \emph{Annals of Physics}, 207:\penalty0 53--70, 1991.
\newblock \doi{https://doi.org/10.1016/0003-4916(91)90178-B}.

\bibitem[Markopoulou and Smolin(2004)]{Markopoulou:2004aa}
Fotini Markopoulou and Lee Smolin.
\newblock Quantum theory from quantum gravity.
\newblock \emph{Physical Review D}, 70:\penalty0 124029, 2004.
\newblock \doi{https://doi.org/10.1103/PhysRevD.70.124029}.

\bibitem[Markopoulou and Smolin(2007)]{Markopoulou:2007aa}
Fotini Markopoulou and Lee Smolin.
\newblock Disordered locality in loop quantum gravity states.
\newblock \emph{Classical and Quantum Gravity}, 24:\penalty0 3813--3824, 2007.
\newblock \doi{https://doi.org/10.1088/0264-9381/24/15/003}.

\bibitem[Pinto-Neto and Struyve(2018)]{Pinto-Neto:2018aa}
Nelson Pinto-Neto and Ward Struyve.
\newblock Bohmian quantum gravity and cosmology.
\newblock https://arxiv.org/abs/1801.03353, 2018.

\bibitem[Roy and Singh(1990)]{Roy:1990aa}
S.M. Roy and Virendra Singh.
\newblock Generalized beable quantum field theory.
\newblock \emph{Physics Letters B}, 234:\penalty0 117--120, 1990.
\newblock \doi{https://doi.org/10.1016/0370-2693(90)92013-9}.

\bibitem[Smolin(2005)]{Smolin:2002aa}
Lee Smolin.
\newblock Matrix models as non-local hidden variables theories.
\newblock In Avshalom~C. Elitzur, S.~Dolev, and N.~Kolenda, editors, \emph{Quo
  Vadis Quantum Mechanics?}, Frontiers Collection. Springer, Berlin, 2005.
\newblock \doi{https://doi.org/10.1007/3-540-26669-0_10}.
\newblock hep-th/0201031.

\bibitem[Smolin(2015)]{Smolin:2015aa}
Lee Smolin.
\newblock Non-local beables.
\newblock \emph{International Journal of Quantum Foundations}, 1:\penalty0
  100--106, 2015.

\bibitem[Starodubtsev(2003)]{Starodubtsev:2003aa}
Artem Starodubtsev.
\newblock A note on quantization of matrix models.
\newblock \emph{Nuclear Physics B}, 674:\penalty0 533--552, 2003.
\newblock \doi{https://doi.org/10.1016/j.nuclphysb.2003.09.053}.

\bibitem[Tilloy(2017)]{Tilloy:2017aa}
Antoine Tilloy.
\newblock Interacting quantum field theories as relativistic statistical field
  theories of local beables.
\newblock https://arxiv.org/abs/1702.06325, 2017.

\bibitem[Unruh(1976)]{Unruh:1976aa}
William~G. Unruh.
\newblock Notes on {B}lack {H}ole {E}vaporation.
\newblock \emph{Physical Review D}, 14:\penalty0 870, 1976.
\newblock \doi{https://doi.org/10.1103/PhysRevD.14.870}.

\bibitem[Vilenkin(1989)]{Vilenkin:1989aa}
Alexander Vilenkin.
\newblock Interpretation of the wave functions of the {U}niverse.
\newblock \emph{Physical Review D}, 39:\penalty0 1116, 1989.
\newblock \doi{https://doi.org/10.1103/PhysRevD.39.1116}.

\bibitem[Vilenkin(1994)]{Vilenkin:1994aa}
Alexander Vilenkin.
\newblock Approaches to quantum cosmology.
\newblock \emph{Physical Review D}, 50:\penalty0 2581--2594, 1994.
\newblock \doi{https://doi.org/10.1103/PhysRevD.50.2581}.

\bibitem[Wald(1994)]{Wald:1994aa}
Robert~M. Wald.
\newblock \emph{Quantum {F}ield {T}heory in {C}urved {S}pacetime and {B}lack
  {H}ole {T}hermodynamics}.
\newblock Chicago Lectures in Physics. University of Chicago Press, Chicago,
  1994.

\end{thebibliography}

\end{document}